\documentclass[sigconf]{acmart}
\usepackage{array}
\AtBeginDocument{%
  \providecommand\BibTeX{{%
    \normalfont B\kern-0.5em{\scshape i\kern-0.25em b}\kern-0.8em\TeX}}}

\setcopyright{acmcopyright}
\copyrightyear{2021}
\acmYear{2021}
\acmDOI{10.1145/1122445.1122456}

\acmPrice{15.00}
\acmISBN{978-1-4503-XXXX-X/18/06}

\begin{document}

    \title{Different Length, Different Needs: Qualitative Analysis of Threads in Online Health Communities}

\author{Daniel Diethei}
\email{diethei@uni-bremen.de}
\affiliation{%
  \institution{University of Bremen}
  \city{Bremen}
  \country{Germany}
}

\author{Ashley Colley}
\email{ashleycolley@gmail.com}
\affiliation{%
  \institution{University of Lapland}
  \city{Rovaniemi}
  \country{Finland}
}

\author{Julian Wienert}
\email{dr.julian.wienert@gmail.com}
\affiliation{%
  \institution{IUBH University of Applied Sciences}
  \city{Bad Reichenhall}
  \country{Germany}
}

\author{Johannes Schöning}
\email{schoening@uni-bremen.de}
\affiliation{%
  \institution{University of Bremen}
  \city{Bremen}
  \country{Germany}
}

\renewcommand{\shortauthors}{Diethei et al.}

\begin{abstract}
Online health communities provide a knowledge exchange platform for a wide range of diseases and health conditions. Informational and emotional support helps forum participants orient around health issues beyond in-person doctor visits. So far, little is known about the relation between the level of participation and participants' contributions in online health communities. To gain insights on the issue, we analyzed 456 posts in 56 threads from the Dermatology sub-forum of an online health community. While low participation threads ('short threads') revolved around solving an individual's health issue through diagnosis suggestions and medical advice, participants in high participation threads ('long threads') built collective knowledge and a sense of community, typically discussing chronic and rare conditions that medical professionals were unfamiliar with or could not treat effectively. Our results suggest that in short threads an individual’s health issue is addressed, while in long threads, sub-communities about specific rare and chronic diseases emerge. This has implications for the user interface design of health forums, which could be developed to better support community building elements, even in short threads.
\end{abstract}




\maketitle
\section{Introduction \& Motivation}
The online discussion forum format includes a range of features that may be of particular benefit to those seeking health information online. Discussion is known to facilitate better learning and knowledge absorption~\cite{laurillard2013rethinking}. This has been identified as a benefit for discussion forums in general~\cite{rowntree1995teaching} and specifically for online discussion forums~\cite{thomas2002learning}. 

In this paper we follow the definition of "online community" from Hammond (2017), who state that online communities are constituted by people who meet together in order to address instrumental, affective goals and at times to create joint artefacts. Interaction between members is mediated by internet technology. In order to constitute community, members need to: show commitment to others; experience a sense of connection (e.g., members need to identify themselves as members); exhibit reciprocity (e.g., the rights of other members are recognised); develop observable, sustained patterns of interaction with others; and show the necessary agency to maintain and develop interaction \cite{hammond2017online}. 

In the domain of social platforms, Jaimes et al. (2011) have defined engagement as the phenomena of being captivated and motivated, and suggest that engagement can be measured in terms of a single interactive session or of a more long- term relationship with the social platform across multiple interactions. Thus, social media engagement is not just about how a single interaction unfolds, but about how and why people develop a relationship with a platform or service and integrate it into their lives \cite{jaimes2011first}. Beuchot \& Bullen~\cite{beuchot2005interaction} as well as Chalkiti \& Sigala~\cite{chalkiti2008information} stated that online forums can promote the collaborative building of information and the management of knowledge. In an analysis of a forum for self-injury support, it was reported that forums construct their own hierarchies. A forum with less moderation assumes a flatter hierarchy where participants rely on each other to set norms of behavior, and if individuals post in a forum in a way that is in conflict with the forum community, other participants will correct the behavior~\cite{smithson2011membership}. 

Shaul~\cite{shaul2008assessing} suggested that online forums could serve as a socially productive learning tool. He divided online forums into three categories: Social/opinion forums, general discussion forums and subject-specific forums. A certain type of subject-specific forums that especially address health-related issues can be labeled as online health communities (OHC). Here, participants seek information related to a specific disease~\cite{dickerson2004patient}, including information that will help them to diagnose a particular health problem~\cite{ybarra2006help}, but also to receive treatment suggestions, often without or before consulting a doctor. Additionally, such OHCs also provide personal experiences of those with a similar condition and social support to participants who seek advice regarding diagnosis, therapy, medication or further strategies. Such communities form a special part of the online support for patients seeking advice or information. Hence, Adler and Adler reported that online discussion forums can go beyond the limits of simple knowledge exploration and can be a source of support for an individual~\cite{adler2008cyber}. 


Another aspect, driven by the increasing load on medical services, is that of supporting individuals to shift at least some of their health information seeking behavior from a face-to-face consultation with a qualified medical practitioner to seeking information online, both before and after diagnosis~\cite{cole2016health}. However, this also requires that the information sourced online is of a reasonably high quality, so as not to pose a health risk. Although previous research on online health knowledge has shown that it is of varying quality ~\cite{impicciatore1997reliability}, a study by Cole et al.~\cite{cole2016health} indicated that forums which contained the most inaccurate or controversial information also contained counterbalancing comments which had the potential to reduce the harmful consequences of poor quality information. Comments made by the original poster, and most respondents in a thread, suggested that the more accurate information carried more influence.



Informed by the prior work, we aim to provide increased understanding of the community building processes at work in online health forums. With this new knowledge we believe that there is potential for the user interface design of such forums to be developed to nurture informative and supportive communities. We address this through the following research questions:
\begin{itemize}
\item RQ1: When do sub-communities emerge in online health communities?
\item RQ2: How do the psycho-social processes vary between threads of different lengths?
\end{itemize}

We report on the detailed analysis of 56 threads from the Dermatology sub-forum of a online health community. In our initial exploration, we identified thread length as a potential criterion for community involvement and separated short (fewer than 26 posts) and long (26 or more posts) threads for analysis.
We found that the issues addressed in the analyzed threads often filled a knowledge or support gap that could not be satisfied by in-person doctor visits. Comparing the level of participation, our results suggest that short threads were helpful for solving an individual's health issues, while in the long threads sub-communities about specific rare and chronic diseases emerged. We propose that the designers of online health community (OHC) user interfaces should aim to better support the collective sensemaking processes present in long threads, e.g., supporting finding information in a thread that was initiated many years ago and grouping related content, making community knowledge accessible even in short threads.

\section{Related Work}
To position our work, we review prior works addressing the use of online medical information in general. We then explore related work that reports on the use of online health communities (OHCs), aiming to identify motivations for their use and types of contributions. 

\subsection{Online Medical Information}
Rather than consulting medical professionals, either in a traditional physical consultation or through a telemedicine service, `Googling symptoms', i.e. seeking self-diagnosis and treatment guidance from freely available online information sources is a common alternative today~\cite{lupton2018digital}. 
Koopman et al.~\cite{koopman2014development} explored patient readiness to engage in health information technology, identifying individual patient's preferred mode of interaction as a key criteria.
Despite individual patient preferences in communication modes, the use of internet self diagnosis has been noted as a source of conflict between patients and medical professionals, by reducing satisfaction with medical professionals when they are later consulted ~\cite{robertson2014my,sjostrom2019primary,lupton2015s}. The main reasons for conflict being due to patients' inability to manage internet sourced information and its potential inaccuracy ~\cite{sjostrom2019primary}. Prior studies about searching online health information have shown that information is of variable quality~\cite{lawrentschuk2012oncology,whitelaw2014internet}. In a study on online health information-seeking behavior, Cole et al.~\cite{cole2016health} found that while a small amount of information was assessed as poor, the original questioner probably would not have been led to act inappropriately based on the information presented. This suggests that online discussion forums may be a useful platform through which people can ask health-related questions and receive answers of acceptable quality.

\subsection{Online Health Communities}

In addition to simply `Googling symptoms', many online discussion forums exist where patients can actively seek community health advice, e.g., reddit's r/AskDocs\footnote{https://https://www.reddit.com/r/AskDocs/} and patient.info\footnote{https://patient.info/forums}. In such forums, online health communities (OHC) can emerge. Besides providing information or suggestions on diagnosis and treatment, OHCs  have also been shown to provide social support among patients with cancer~\cite{setoyama2009peer}, diabetes~\cite{huh2012collaborative}, rare diseases~\cite{lasker2005role}, infertility~\cite{welbourne2009supportive}, and HIV/AIDS~\cite{mo2010living}. Previous studies described OHCs members’ information seeking practices~\cite{deChoudhury2014seeking}, as well as ways they access and appraise information~\cite{hartzler2014evaluating,huh2012collaborative}, and construct new knowledge together~\cite{mamykina2015collective}. Based on analysis of a diabetes forum, Mamykina et al.~\cite{mamykina2015collective} describe the value of the forum as exposing individuals to the richness and multiplicity of different perspectives, which help participants to construct their own personal views. 

\subsubsection{Motivations}
To understand the behavior of users in online health communities, it is important to know why they visit and join such platforms. Huh~\cite{huh2015clinical} identified themes in an online diabetes forum, suggesting that thread initiators' motivations were often not only to get clinical expertise, but also to hear other patients’ personal experiences. Other motivations for posting in the forum included not wanting to see a doctor, wanting immediate discussion before a scheduled doctor's appointment, or a second opinion following a doctor's diagnosis~\cite{huh2015clinical}. Investigating the motivations of pregnant women for seeking support online, Gui et. al~\cite{gui2017invest} identified limited access to healthcare professionals, frustration with their healthcare providers, limited access to offline support and a mismatch between information from books of internet resources with their own experience as main factors. The types of support sought by pregnant women included advice, formal and informal pregnancy-related knowledge, reassurance and emotional support. The posts mainly addressed sharing experiential knowledge, passing on other healthcare providers' opinions, suggesting going for professional help and action based on peers' responses~\cite{gui2017invest}. Tied to participants' motivations are social roles which vary among OHC members. Yang et al.~\cite{yang2019seekers} identified 11 roles in an online cancer support community three of which are emotional support provider, informational support provider and story sharer. The authors found that the roles predict long-term participation and change over time from ones that seek resources to ones offering help.

\subsubsection{Types of Contributions to Discussions in OHCs}
Mamykina et al.~\cite{mamykina2015collective} identified ten different types of contribution posts make to the discussion in a health forum, including asking a question, suggesting resources, agreement/disagreement with a previously stated position, further developing previously stated position, personal reconciliation and synthesis of previously stated perspectives. 
Besides informational support, OHCs provide emotional support, often defined as a response of the community to a member’s desire to change their mental state, usually to be more optimistic, motivated and determined~\cite{nambisan2011information}. This support can take many forms but usually requires that a member who seeks it is integrated with the community and can capitalize on the existing social structures~\cite{house1988social}. Building on these results, Nakikj \& Mamykina~\cite{nakikj2017park} conceptualize socio-emotional needs as a member’s desire to change their mental state through social interaction with other members and as a result of their social integration with the group.
    
In summary, previous research established that patients frequently seek health information online. They consult online health communities not only for informational but also for psycho-social support. Prior work also highlighted the motivations that drive members of online health communities. In addition, past research addressed the types of contributions common in such communities. We aim to extend existing literature by juxtaposing the different mechanisms of engagement in short and long threads in OHCs and by exploring the collective knowledge construction and community building.

\section{Method}

In this study, we conducted a qualitative analysis of a sub-forum in a large English-speaking online health community. Although all the information posted to the forum is publicly available,  we do not name the specific community to avoid reverse identification of participants. The forum contains sub-forums addressing many medical specialties. Threads contain text and optionally images. Responses can be posted to up to three levels, e.g. an original post can have up to two levels of subthreads.

To familiarize ourselves with our area of study, we first studied the ``DermatologyQuestions" sub-forum on Reddit (so called subreddits, e.g. r\textbackslash Dermatology). Reddit is a discussion website that hosts a range of non-medical and medical subreddits. Since the ``DermatologyQuestions" consists of a large community, including some medical professionals, we first conducted an online interview with a community member, who was also a certified dermatologist. We then developed a thematic codebook by analyzing 150 threads (including 244 posts) from r\textbackslash DermatologyQuestions. Analysis revealed that the Reddit posts often comprised only of short questions and answers, making it difficult to identify types of contributions and community dynamics. 

\begin{figure}
\centering
\includegraphics[width=0.98\columnwidth]{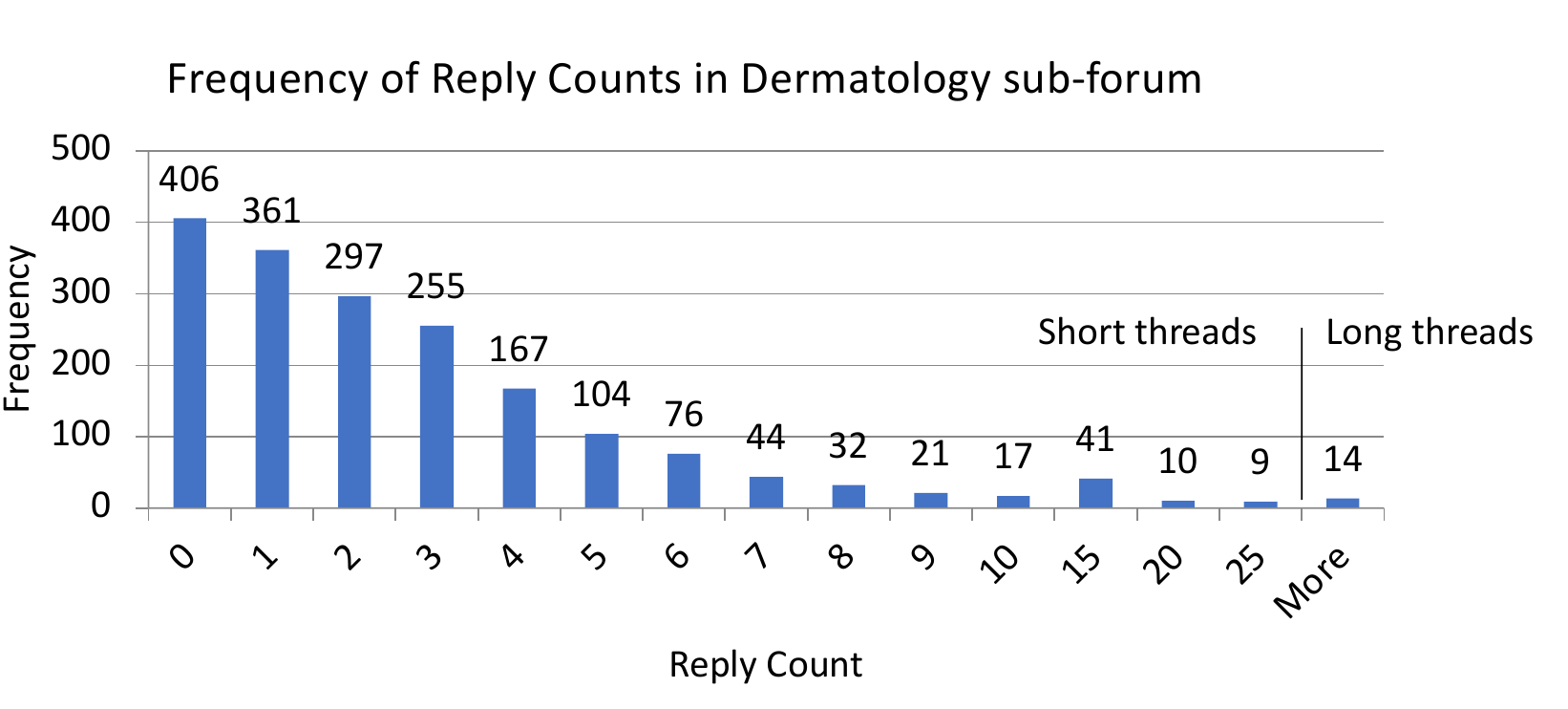}
\caption{Frequency of reply counts of threads (n=1854) in the Dermatology sub-forum on the online health community (specific platform not mentioned to ensure participant privacy). The majority of threads (n=406) has no reply. Counts for threads with 11-15 (15), 16-20 (20), 21-25 (25) posts are aggregated into the respective category. Included are all threads publicly available as of January 27th, 2021. We selected the most recent 50 threads with fewer than 26 posts and the most recent six threads with 26 or more posts for our study.}
\label{fig:post_count}
\end{figure}

With our developed codebook, we shifted our focus to analyse another site that included health forums which appeared to have richer discussions. We specifically focused on the sub-forum ``Dermatology", as our initial exploration revealed that it was among the most popular sub-forums and included threads with a variety of lengths. For the main analysis of this paper, we used our developed codebook to analyze 50 short (with a total of 124 posts) and six long threads (with a total of 332 posts) from the OHC. The frequency of various thread lengths in the OHC is visualised in Figure \ref{fig:post_count}.

\subsection{Data Collection}

Our dataset consists of textual information and images that were directly uploaded to the post or linked on image hosting websites, e.g., imgur.com. 
To retrieve the OHC posts, we used the python package \textit{beautifulsoup} to parse the source code. All data was collected between July 29 and August 27, 2020.

\subsection{Analysis}

We applied open coding combined with thematic analysis in line with Blandford et al.~\cite{blandford_qualitative_2016}. For the analysis of the data we created a jupyter notebook with the python package \textit{ipyannotate} with some modifications for the forum data hierarchy and to display the photos together with the post. Existing software for qualitative data analysis was not considered suitable for this rather specific task. For the preparation of the codebook, author DD developed an initial codeset based on prior work on types of contributions to online health communities~\cite{mamykina2015collective}. The codeset was then refined and extended with authors AC and JW, e.g. we adopted the code ‘asking question’ but also introduced new codes (i.e., ‘giving advice’ and ‘own experience’). Moreover, we added symptoms, such as pain, and body parts to the list of categories. 

As our initial analysis revealed that some longer threads span over multiple years and seemed to be of high significance for a group of participants with similar health problems, we decided to investigate thread length as a factor. We initially identified a small number of threads that contained a large amount of posts. Through review of these threads, we noted one thread, \textit{moisturizing lotion}, that was just long enough that group building elements were beginning to emerge.  Hence, we defined the minimum length of what we refer to as long threads, based on the length of this thread. We are aware that this is a somewhat arbitrary definition. For our main study, we analyzed 50 short threads, i.e. with fewer than 26 posts, and six long threads, with 26 or more posts, in total 332 posts. We iteratively compared our coding tree with our dataset until we reached data saturation \cite{blandford_qualitative_2016}. All the analyzed threads are still open and accepting new posts.

\begin{figure*}
\centering
\includegraphics[width=\textwidth]{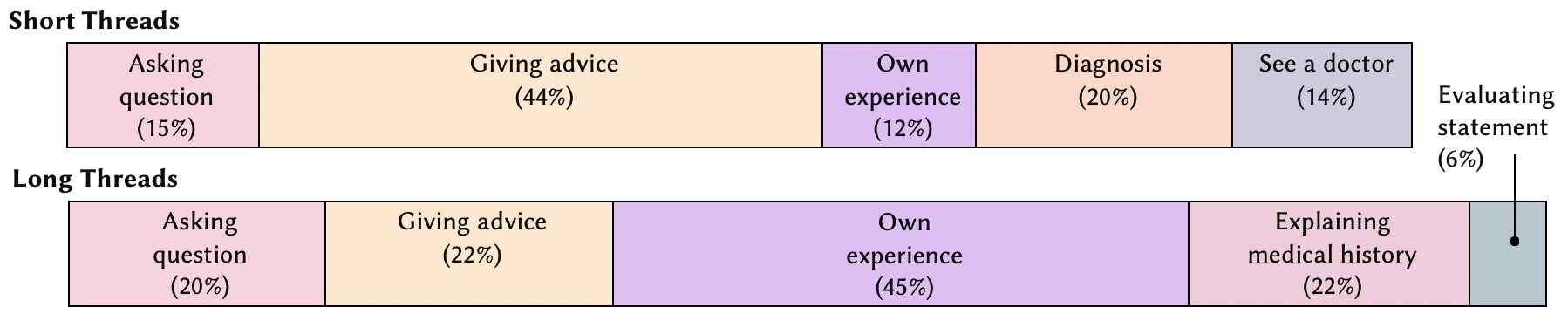}
\caption{Comparison of five most frequent themes in the short (n=50, 124 posts) and the long (n=6, 332 posts) threads. While the posts in the short threads were mostly direct support through giving advice, e.g. about medications and treatments, members in the long threads presented their own experience with their medical condition most frequently. Often, posts in the short threads included diagnosis suggestions, which was not present in the long threads. One post could be assigned multiple themes, therefore the percentages do not add up to 100.}
\label{fig:codes_frequency}

\end{figure*}

\section{Results}
In the following section we first describe key differences between short and long threads, then, secondly, present the qualitative content analysis, and, thirdly, explain our model of engagement in long threads..
To ensure participants' privacy, we do not mention user names when citing forum posts.

\begin{table}[b]
\centering
\begin{tabular}{ m{1.4cm} | m{0.6cm}| m{1.2cm}| m{1.3cm} | m{1.3cm} | m{0.7cm}} 
\textbf{Thread} & \textbf{Posts} & \multicolumn{2}{c|}{\textbf{Post Length (words)}} & \textbf{Time Span} & \textbf{Partici-pants}\\
\hline
Cat Scratches & 128 & \textit{M} = 91.5 & \textit{SD} = 110.5 & 5 years, 10 months & 47\\ \hline
Pityriasis Versicolor & 42 & \textit{M} = 98.8 & \textit{SD} = 66.1 & 5 years, 4 months & 16 \\ \hline
Cannot sweat & 35 & \textit{M} = 131.0 & \textit{SD} = 106.9 & 4 years & 13\\ \hline
Moisturising  lotion & 26 & \textit{M} = 50.8 & \textit{SD} = 47.3 & 5 days & 4\\ \hline
PLC sufferer & 87 & \textit{M} = 104.1 & \textit{SD} = 94.3 & 6 years, 4 months & 39\\ \hline
Sebaceous cysts & 50 & \textit{M} = 153.5 & \textit{SD} = 124.2 & 11 years, 5 months & 32\\ \hline
\end{tabular}
\caption{Descriptive statistics of long threads. For "Participants", all posts made by guests were counted as one participant.}~\label{tab:stats}
\end{table}

\begin{table*}[t]
\centering
\begin{tabular}{ m{2cm} |  m{14cm}} 
\textbf{Thread} & \textbf{Summary}\\
\hline
Cat Scratches & This condition is mystery to medical professionals, who don't believe the conditions exists. Diagnoses include fungal, parasites, connection to Lyme disease. Potential causes are ambient temperature, household mold. Treatments suggested are ointments, camphor oil, bathing in vinegar, covering the body with a paste of coconut oil and colloidal silver. \\ \hline
Pityriasis Versicolor & Most participants have identified the disease they are suffering from (Pityriasis Versicolor). Participants share treatments, e.g. creams, anti-fungal medication, medicated shampoo, coconut oil. Also dietary changes (low sugar, gluten free, dairy free and mushroom free) are suggested. Symptoms are reported as particularly affecting the mental health of the participants. \\ \hline
Cannot sweat & Practices to mitigate the painful effects of the condition of being unable to sweat. The original poster (OP) posts the diagnosis of his condition (Cholinergic Urticaria), noting that it has no permanent cure, but proposes an treatment regime (physical exercise) that has reduced the reduce the severity of the symptoms by 80\%. For several of the participants the original poster has become the expert on the condition.\\ \hline
PLC sufferer & Participants in this thread suffer from the condition pityriasis lichenoides chronica (PLC), which is characterized as rare and mostly unexplored. Many have had the condition for more than 10 years. The thread discussion concludes in a treatment regime that has worked for several participants. Due to the lack of a cure, participants report mental challenges. \\ \hline
Sebaceous cysts & Participants in this thread suffer from a condition of reoccurring cysts on the bikini line. Many report having the condition for more than 10 years, undergoing multiple painful surgical procedures and accruing \$1000 in medical bills. Participants express frustration with medical professionals. A strong theme is that the forum provides the feeling of not being alone with the condition.  \\ \hline
Moisturing Lotion & Participants discuss allergic reactions to certain brands of moisturizing lotion, providing details on the products they are using and highlighting the ingredients that are the potential cause of the reaction. This is a relatively recent thread, and, at time of writing, 7 months have passed since the last post. Only 5 participants take part in the thread with more than half of the posts (14/26) being made by the OP. The discussion is particularly light-hearted and chatty, e.g., when discussing a possible allergy to formaldehyde the OP jokes about the shock the undertaker will get when they are embalmed. \\ \hline
\end{tabular}
\caption{Content summaries of the six long threads.}~\label{tab:summaries}
\end{table*}

\subsection{Comparison between Short and Long Threads}
Figure \ref{fig:codes_frequency} presents the five most frequent themes in the analyzed short (n=50, 124 posts) and long (n=6, 332 posts) threads. Whilst some similarities are apparent, there are clear differences in the weighting of themes between the two cases and several themes, e.g. diagnosis and see a doctor, were only identified in one thread type.

The initial posts in the analyzed short and long threads were generally similar, describing the symptoms with which the poster was suffering. However, in the case of short threads, a diagnosis was often specifically sought (present in 20\% of posts), e.g. `What is this?'. In almost all the analyzed long threads the participants had a clear understanding of the medical condition with which they were suffering. Here, the \textit{Cat scratches} thread was an exceptional case where there was debate on the underlying cause of the symptoms. The content of the long threads is summarized in Table \ref{tab:summaries} and descriptive statistics are shown in Table \ref{tab:stats}.

A clear difference between short and long threads was the difference in balance between sharing one's own experiences (12\% vs. 45\%) and giving advice (44\% vs. 22\%). In short threads, the discussion was mainly focused towards solving a health issue of one specific user. Respondents were typically not suffering from the health issue themselves, but provided advice, e.g., treatment strategies and medications, or shared diagnosis suggestions. Sometimes (14\% of posts), posters were instructed to see a  doctor if respondents suspected potentially harmful medical issues or were unsure how to help.

Long threads revolved around a single health issue that was shared by all the participants. Altogether, 67\% of posts included descriptions of personal medical history (22\%) or experiences (45\%), sometimes together with advice, e.g. reporting on cysts after cycling long distance and recommending laser hair removal and hydrocortisone cream.
Almost without exception, respondents were affected by the same condition, therefore there was less discrepancy in the role participants played in the discussion. As in most cases participants had already seen, often multiple, medical professionals with unsatisfactory outcomes, there was very rarely any advice given to consult a doctor. 

Analysis of the length of individual posts in both thread types revealed a significant difference in the length of post for short threads (\textit{M} = 75 words, \textit{SD} = 84) vs. long threads (\textit{M} = 102 words, \textit{SD} = 104); t(558) = -3.38, p \textless .001. Approximately 7\% of posts in short threads contained more than 200 words, whilst for long threads this amount was double (14\%). Differences between the analyzed long threads are also notable. In the \textit{Sebaceous cysts} thread almost 30\% of the posts exceeded 200 words, whilst in the \textit{Moisturising lotion} case the longest post was 124 words in length.
\vspace{3cm}



\subsection{Model for Community Building in Online Health Communities}
\begin{figure*}
\centering
\includegraphics[width=\textwidth]{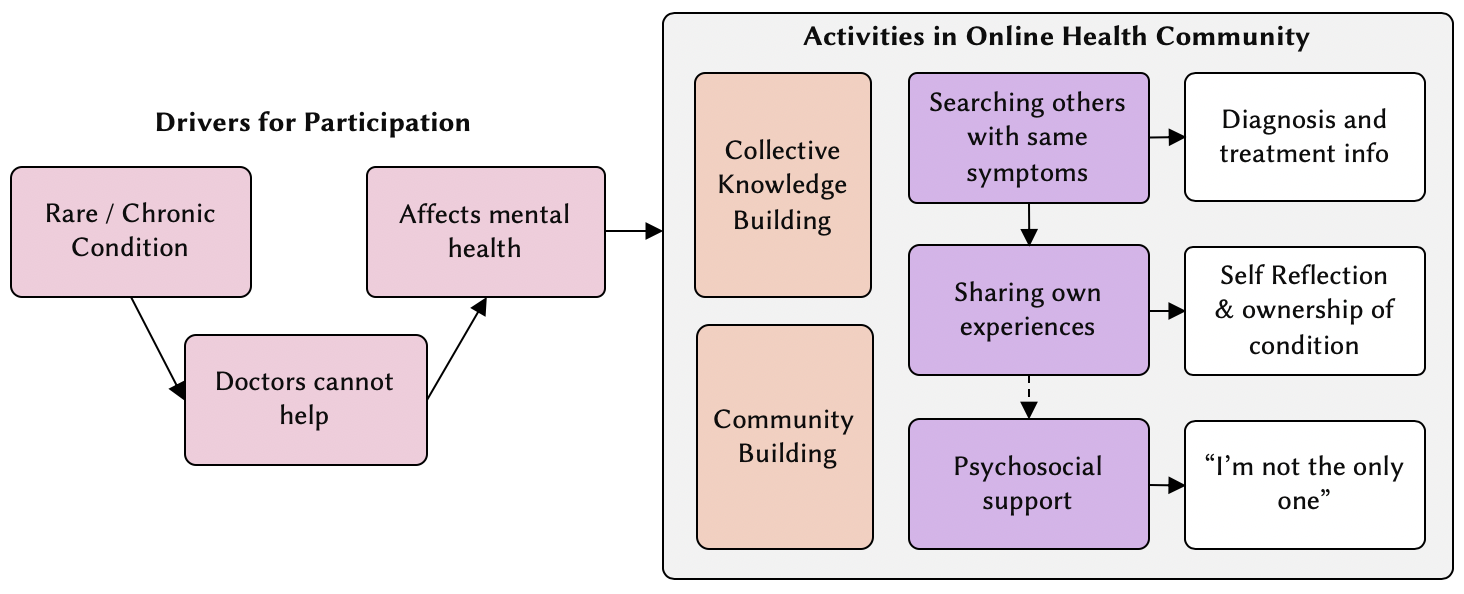}
\caption{Model of high engagement threads in online health communities, that lead to the development of sub-communities}
\label{fig:model}
\end{figure*}

Based on our findings from exploring the novel characteristics of long threads, we formulated a model of engagement (see Figure~\ref{fig:model}) to explain the psycho-sociological processes of collective knowledge building and community building that we observed in such threads. The driver of an individual to participate in such threads is suffering from a rare or chronic diseases that likely affects their mental health. When doctors can not help, the mental condition may further deteriorate. Patients therefore seek help in online health communities and actively search for other people with the same symptoms they suffer from. Many will share their own experiences, adding to the knowledge pool of the community and providing a tool to support their own self reflection. This exchange of experience and psycho-social support, often simply relief that there are others with the same condition, creates a sense of community among the participants.

Reflecting on the six long threads in our dataset against the identified criteria for community building, we notice that the \textit{Moisturising cream} thread, whilst meeting our thread length threshold, does not support any of the other criteria defined in our model of engagement. The described topic is neither rare or chronic and rather than negatively affecting mental health, the discussion in the thread is of a jovial nature. It is also notable that the length of posts in the thread is far below the average identified for the other five community building long threads (see Table \ref{tab:stats}).

\subsubsection{Drivers for Participation}
Community building threads emerge around participants' experiences of conditions that affect them severely and for long periods of time. Some participants in our dataset reported a history of having suffered with conditions for more than 20 years, including frequent doctor changes and unsuccessful treatments. 
 
Analysis of the long threads revealed two reasons of why participants looked for advice beyond that from medical professionals. Firstly, doctors were not able to help. Participants reported being prescribed a multitude of different creams, tablets, management regimes and even undergoing repeated painful and costly operations -- often with little or no success. Secondly, participants reported that doctors were often unfamiliar with their conditions, possibly because the conditions addressed in the threads were typically rare.

It was also apparent from the long threads that many participants had made extensive research on their condition, and, based on this, critically evaluated any advice they received from doctors. In some cases, participants reported dissatisfaction with their doctors, considering the doctor had not done enough research or describing the doctor's behavior as rude. Some participants proposed showing the discussion from the thread to their doctor, e.g., reporting that they would bring posts from this forum to their dermatologist so that they would understand that these are not scratch marks and many people are going through this.
Another participant highlighted the frustration of doctors not considering their condition to be real, explaining that they had been banned from the dermatology department at their local hospital and doctors suggest the participant is imagining it.
Further, in some cases the treatment proposed by some doctors was considered impractical, e.g. one participant reported she was instructed to stay out of the sun, which she considered bad advice as she is young and wants to go on holidays.
Hence, alternative solutions were sought through own research and exchange in the forum. 

The severity and chronic nature of the symptoms led participants to frequent comments on their mental health. Participants did not know what to do next, resulting in feelings of hopelessness, frustration and desperation. 
For example participants stated that some symptoms are so slow developing that it feels like nothing has changed until one day they realized how gray everything feels. Another participant resonated that she gets depressed while breast feeding her baby and that she is struggling with PLC and, having had the disease for 8 years, she feels hopeless.  

The dermatological conditions in the scope of our study affected patients' quality of life in multiple ways. For example, efforts to maintain high hygiene, both in the domestic environment and the body, impacted to daily routines, e.g. steam cleaning bedding and washing hair daily. 
Similarly, participants reported needing to wear clothes that covered the parts of their skin affected by the condition. Feelings of shame were therefore associated with exposing the condition in public. Other participants had to stop leisure activities such as bike riding, due to cysts appearing in the groin area.

\subsection{Activities in Online Health Community}
The main value of the long thread discussions was the building of collective knowledge and community among the thread participants. From our dataset, we identified a strong motive in the long threads was the exchange of informational and emotional support. Participants encouraged the community to contribute their experiences to detect patterns, identify successful strategies and spread knowledge. Participants shared their experiences primarily through lengthy detailed descriptions of their medical history. When a successful strategy was shared, other participants followed the advice. There was active exchange on the strategies, often resulting in participants demanding updates on others' proposed ideas, e.g., following up to ask whether they had tried certain medications.
Consequently, progress reports of successful therapies were also shared, e.g. reporting that their PLC is 75\% gone, having eaten healthy for seven weeks, which consequently changed the quality of their life and also encouraging others to beat the disease together.

\begin{figure*}
\centering
\includegraphics[width=\textwidth]{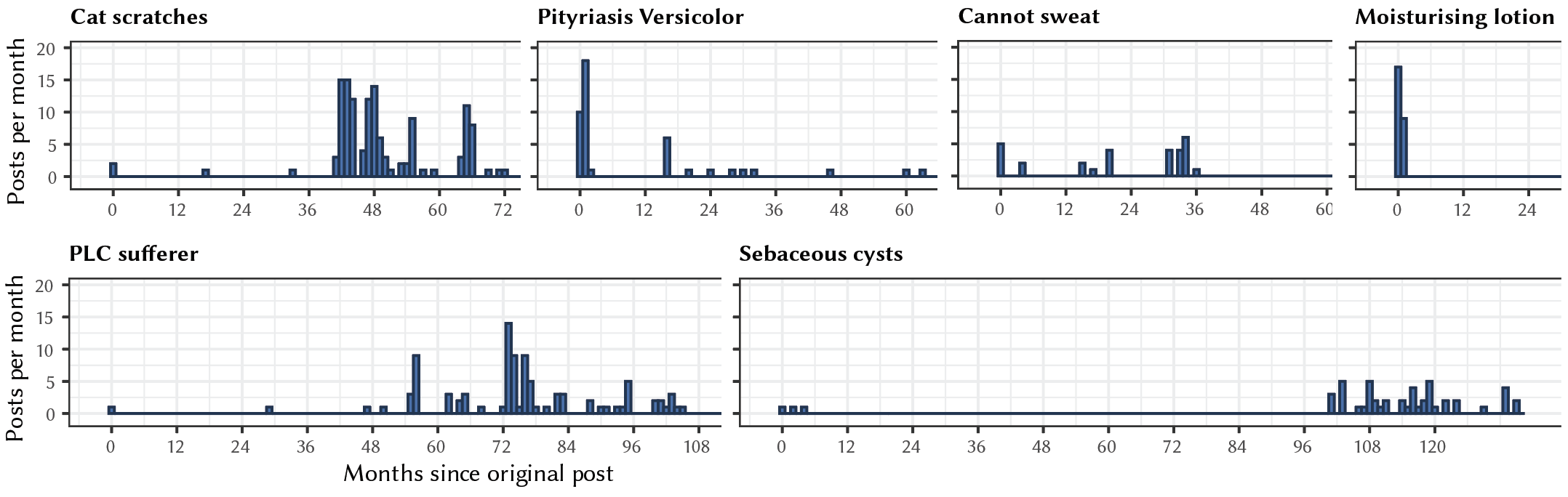}
\caption{Timeline of posts in each of the analyzed long threads (* For cat scratches and Pityriasis Versicolor the original post had been later edited, thus the origin time was taken as that of the second post)}
\label{fig:timeline}
\end{figure*}

However, feedback on the success of shared medication suggestions and treatment strategies was often missing.  Participants indicated that they wanted to follow their peers' advice, sometimes even asking follow up questions, e.g. the exact name of a medication, but rarely reported back to the forum on the success of the strategy. Similarly, when members announced that they were planning to visit a doctor, other members asked for updates on the outcome. In some cases information was forthcoming, but in others the thread dried up, even though the information was begged by multiple different participants. 

Support was not limited to exchanging medical advice, participants created a sense of belonging through providing reciprocal psycho-social support. This was expressed as relief when hearing about other participants with similar experiences. While it seemed that participants' doctors, family and friends had trouble understanding the physical and mental implications of their condition, participants found empathy in the forum, e.g. letting other participants know that there is a solution and that they are not alone.
We observed that the sense of community was a key factor in managing the emotional stress associated with the participants' conditions.  

\subsection{Emergence of Sub-Communities}
Active participation in forum threads over long periods of time formed sub-communities around a specific health problem experienced by many participants. The birth of such a sub-community requires a critical mass of participants, evident from the extremely long time period many of the analyzed long threads were dormant, before becoming active (see Figure \ref{fig:timeline}). Some participants stayed active in the community over several years. Interestingly, the notion of time appeared to be of limited importance to the community participants, with many participants responding to posts from more than 2 years previously. Rather, what was important was finding people with the same medical condition. 

From our dataset we noted that patients were often left behind the ``real world'' and found solutions to their health problems exclusively in the community. This applied to emotional support too. Some posters noted that they felt relief having found other people with similar issues. Apparently, the real word social support circles, such as family and friends, could not care for the participants' specific social needs. This gap was mitigated through psycho-social support exchanged in the forums.

\section{Discussion}


From our analysis we find that one of the main motivators for people to search for medical information in long threads is that they are not satisfied with their doctor, either because the doctor could not help, often due to the condition being rare or unexplored, or because the doctor did not take them seriously (RQ1). Similar feelings, as motivation for patients to participate in online health communities, have been reported by Gui et al.~\cite{gui2017invest}. While there were some occasions where forum participants forwarded health information they had received from medical professionals, such as the names of prescribed medications, many discussions revolved around treatment strategies developed through personal experience while suffering from a condition, sometimes over many years or even decades. Here our findings align with those of Huh et al.~\cite{huh2015clinical}, who identified searching for other patient's personal experiences and avoiding doctors as a common motives for consulting online health communities. While we did not analyze the social roles of participants in detail, we noted that in one instance the initiator of the 'Cannot sweat' thread clearly transitioned from seeking information to offering advice, suggesting that motivations can change during the exchange with fellow patients. From these observations we conclude that the discussion forums were a valuable resource for both informational and emotional support and compensated for needs that had not been satisfied outside of the online health communities. While similar findings have been reported by others~\cite{house1988social,nakikj2017park,nambisan2011information,mamykina2015collective}, we found that such discussions in the long threads emerged around rare and chronic diseases when healthcare providers could not provide support anymore.

Our results also suggest that threads with different lengths serve different purposes (RQ2). While short threads are often directed towards an individual's condition, longer threads, i.e. with 26 or more posts, evolved into sub-communities, where a sense of community developed among participants. In the short threads, there was usually a clear distinction of roles between those participants asking for advice and those giving advice. In the long threads, however, participants had similar motivations, i.e. discussing a condition they were all suffering from, and therefore the social roles were more similar. Referring to Beuchot \& Bullen \cite{beuchot2005interaction}, who report that forums build their own hierarchies, we note that longer threads had a flatter hierarchy 
than short threads which had a deeper information seeker and advice giver hierarchy. 

\subsection{Implications}
Participants who access long threads many years after the thread discussion started may have difficulty identifying relevant health advice. These threads sometimes span over dozens of posts across many years, making it hard for participants to find advice that has been ``validated" by the community. Additionally, in long threads discussions might become divided into multiple streams, i.e. replies to posts, and discussions in parallel and detached from each other. We suggest that OHC developers should account for this and, for example, provide a summary of posts that include health advice that has been found useful by the community, e.g. successful treatment strategies. 

Further, we discovered that threads with more participation led to deeper involvement of individuals with their condition and the community. Thus, our proposals to provide summaries of threads and highlight helpful posts could be extended to a sub-forum level, allowing for a intelligent search function based on medical categories, e.g. diagnosis and treatment. Threads and posts in sub-forums could be automatically grouped into categories and participants would then enter a keyword into the search bar, e.g. rash, to navigate through different threads that address the same or similar medical conditions. With this approach, even short threads would become embedded into a larger context, supporting knowledge construction for individuals seeking health advice. 


\subsection{Limitations}
We acknowledge that our results are limited in their generalisability, as we collected our data from a single sub-forum focusing solely on dermatological conditions. By sampling across various communities and sub-forums (e.g., other medical domains), future work could extend the applicability of our findings. Nevertheless, we believe our current findings contribute an important first step in exploring approaches to provide better medical support via OHCs.

\section{Conclusion}
Based on the analysis of a dermatology focused OHC, we identified that the number of posts in a thread is indicative of the purpose of the thread, the level of collective sensemaking processes as well as the duration of participation. While short threads addressed health conditions of an individual and were mostly in a question and answer format, long threads, with more than 26 posts brought forward sub-communities where participants shared own experiences and provided mutual informational and emotional support. We provide suggestions for developers of OHCs to allow participants to take advantage of some community building and knowledge construction elements even in threads with low participation.


\bibliographystyle{ACM-Reference-Format}
\bibliography{refs}

\end{document}